\documentclass[twocolumn,showpacs,prb,superscriptaddress,aps,floatfix]{revtex4-1}
\usepackage{rotating}
\usepackage{amsmath}
\usepackage{color}
\usepackage{graphicx}
\usepackage{epsfig}
\usepackage{mathrsfs}
\usepackage{courier}
\usepackage[sort&compress]{natbib}

\newcommand{\lv}{{\bf a}}
\newcommand{\bb}{{\bf b}}

\newcommand{\qq}{{\bf q}}

\newcommand{\rr}{{\bf r}}
\newcommand{\pp}{{\bf p}}
\newcommand{\PP}{{\bf P}}
\newcommand{\kk}{{\bf k}}

\newcommand{\be}{\begin{equation}}
\newcommand{\ee}{\end{equation}}
\newcommand{\ben}{\begin{equation*}}
\newcommand{\een}{\end{equation*}}
\newcommand{\bea}{\begin{eqnarray}}
\newcommand{\eea}{\end{eqnarray}}
\newcommand{\bean}{\begin{eqnarray*}}
\newcommand{\eean}{\end{eqnarray*}}

\renewcommand{\[}{\left[}
\renewcommand{\]}{\right]}

\def\efield{\boldsymbol{\cal E}} 
\def\ket#1{\vert#1\rangle}

\def\susc#1{\chi^{(#1)}}
\def\ket#1{\vert#1\rangle}

\def\ai{\emph{ab-initio}\ }

\newcommand{\cnrs}{CNRS/Univ. Grenoble Alpes, Institut N\'eel, F-38042 Grenoble, France} 
\newcommand{\qub}{School of Mathematics and Physics, Queen's University Belfast, Belfast BT7 1NN, Northern Ireland, UK}
\newcommand{\coimbra}{Centre for Computational Physics and Physics Department, University of Coimbra, Rua Larga, 3004-516 Coimbra, Portugal}

\begin{document}
\title{Non-linear optics from an ab-initio approach by means of the dynamical Berry-Phase:\\
application to second and third harmonic generation in semiconductors}
\author{C. Attaccalite}
\affiliation{\cnrs}

\author{M. Gr\"uning} 
\affiliation{\qub}
\affiliation{\coimbra}

\begin{abstract}
We present an \ai real-time based computational approach to study nonlinear optical properties in Condensed Matter systems that is especially suitable for crystalline solids and periodic nanostructures.  The equation of motions and the coupling of the electrons with the external electric field are derived from the Berry phase formulation of the dynamical polarization [Souza et al. Phys. Rev. B 69, 085106 (2004)]. Many-body effects are introduced by adding single-particle operators to the independent-particle Hamiltonian. We add a Hartree operator to account for crystal local effects and a scissor operator to correct the independent particle band structure for quasiparticle effects. We also discuss the possibility of accurately treating excitonic effects by adding a screened Hartree-Fock self-energy operator.
%Correlation effects are included by modifying the zero-field Hamiltonian. In particular the quasi-particule band structure can be used as starting point and crystal local field plus excitonic effects can be included in the response functions. 
The approach is validated by calculating the second-harmonic generation of SiC and AlAs bulk semiconductors: an excellent agreement is obtained with existing \ai calculations from response theory in frequency domain [Luppi et al. Phys. Rev. B 82, 235201 (2010)]. We finally show applications to the second-harmonic generation of CdTe and the third-harmonic generation of Si.

%The zero-field Hamiltonian includes crystal local field effects and the renormalization of the independent particle energy levels by correlation effects. We also discuss the possibility of accurately treating excitonic effects by adding a screened Hartree-Fock self-energy operator.The approach is validated by calculating the second-harmonic generation of SiC and AlAs bulk semiconductors: an excellent agreement is obtained with existing \ai calculations from response theory in frequency domain (Phys. Rev. B 82, 235201). We finally show applications to the second-harmonic generation of CdTe and the third-harmonic generation of Si.
\end{abstract}           

\maketitle

\section{Introduction}
\emph{Ab-initio} approaches based on Green's function theory became a standard tool for quantitative and predictive calculations of linear response optical properties in Condensed Matter. In particular, the state-of-the-art approach combines the $G_0W_0$ approximation for the quasi-particle band structure~\cite{aryasetiawan1998gw} with the Bethe-Salpeter equation in static ladder approximation for the response function.~\cite{strinati} This approach proved to effectively and accurately account for the essential effects beyond independent particle approximation (IPA) in a wide range of electronic systems, including extended systems with strong excitonic effects.~\cite{RevModPhys.74.601}

In contrast, for nonlinear optics \ai calculations of extended systems rely in large part on the IPA\cite{PhysRevB.48.11705} with correlation effects entering at most as a rigid shift of the conduction energy levels.~\cite{PhysRevB.80.155205}  Within time-dependent density-functional theory (TDDFT), it has been recently proposed~\cite{PhysRevB.82.235201} an approach to calculate the second-harmonic generation (SHG) in semiconductors that takes into account as well crystal local-field and excitonic effects. However, this promising approach~\cite{Cazzanelli2012} is limited by the treatment of the electron correlation to systems with weakly bound excitons.~\cite{PhysRevB.69.155112} 

%In fact, it is extremely involved to include many-body effects into the expression for the nonlinear optical susceptibilities within Green's function theory. 
%In fact, the inclusion of many-body effects in the equations for the non-linear susceptibilities makes them very difficult to solve.
Within Green's function theory the inclusion of many-body effects into the expression for the nonlinear optical susceptibilities is extremely difficult. 
Furthermore the complexity of these expressions grows with the perturbation order. Therefore it is not surprising that there have been only few isolated attempts to calculate second-order optical susceptibility using the Bethe-Salpeter equation~\cite{Leitsmann2005,Chang2002} and no attempt to calculate higher-order optical susceptibilities.~\cite{PhysRevB.80.165318} 
%The complexity growing with the perturbation order, there have been only few isolated attempts at calculating second-order optical susceptibility using the Bethe-Salpeter equation,~\cite{Leitsmann2005,Chang2002} and in practice third-order optical susceptibility is untreatable.~\cite{PhysRevB.80.165318} 
%On the other hand, the origin of this limitation is the attempting to calculate the nonlinear optical susceptibility directly in terms of the electronic structure. 

Alternatively to the frequency-domain response-based approach, one can obtain the nonlinear optical susceptibility in time-domain from the dynamical polarization $\PP$ of the system by using the expansion of $\PP$ in power of the applied field
\be
\label{eq:peopbf}
\PP= \chi^{(1)} \efield + \chi^{(2)} \efield^2 + \chi^{(3)} \efield^3 + \dots 
\ee 
This strategy is followed in several real-time implementations of TD-DFT~\cite{PhysRevB.54.4484}. In these approaches the dynamical polarization is obtained by numerical integration of the equations of motion (EOMs) for the Kohn-Sham system.~\cite{takimoto:154114,castro:3425,meng:054110} So far applications regard mostly nonlinear optical properties in molecules.

The time-domain approach presents three major advantages with respect to frequency-domain response-based approaches. First, many-body effects are included easily by adding the corresponding operator to the effective Hamiltonian. Second, it is not perturbative in the external fields and therefore it treats optical susceptibilities at any order without increasing the computational cost and with the only limitation dictated by the machine precision. Third, several non-linear phenomena and thus spectroscopic techniques are described by the same EOMs. For instance, by the superposition of several laser fields one can simulate sum- and difference-frequency harmonic generation, or four-waves mixing.~\cite{boyd}

In a recent work,~\cite{attaccalite} we proposed a real-time implementation of the Bethe-Salpeter equation, based on nonequilibrium Green's function formalism.
However, due to the problems in defining the position operator and thus $\PP$, it is not trivial to apply Eq.~\eqref{eq:peopbf} to systems in which periodic boundary conditions (PBC) are imposed. As it was recognised for example in Ref.~\onlinecite{PhysRevB.52.14636}, the same problem appears in the direct evaluation of the nonlinear optical susceptibility in frequency-response based approaches. In particular the dipole matrix elements between the periodic part of the Bloch functions are ill-defined when using the standard definition of the  position operator. In that case, it is possible to obtain correct expressions for the dipole matrix elements from perturbation theory~\cite{PhysRevB.52.14636,PhysRevB.48.11705,PhysRevB.82.235201} at a given order in the external field. Instead, in the real-time approach one needs an expression valid at each order of the perturbation.

A correct definition of the polarization operator in systems with PBC has been introduced by means of the geometric Berry phase in the Modern theory of polarization.\cite{RevModPhys.66.899} 
To our knowledge real-time schemes for calculating the electron-field coupling consistently with PBC have been proposed in Refs.~\onlinecite{springborg, PhysRevB.76.035213, souza_prb}. In those works the dipole matrix elements are evaluated numerically from the derivative in the crystal-momentum ($\kk$) space. The latter cannot be carried out trivially because of the freedom in the gauge of the periodic part of the Bloch functions. In fact, the gauge freedom leads to spurious phase differences in the Bloch functions at two neighbouring $\kk$ points and ultimately to spurious contributions to the numerical derivative.
Then, basically the three schemes~\cite{springborg, PhysRevB.76.035213, souza_prb} differ in how the gauge is fixed to eliminate the spurious phase.

This work presents a real-time \ai approach to nonlinear optical properties for extended systems with PBC in which the nonlinear optical susceptibility are obtained through Eq.~\eqref{eq:peopbf}. To derive the EOMs we follow the scheme of Souza et al.\cite{souza_prb} based on the generalization of Berry phase to the dynamical polarization (Sec.~\ref{ss:fldcpl}). Originally applied to a simple tight-binding Hamiltonian, this approach is valid for any single-particle Hamiltonian and, as we discuss in Sec.~\ref{ss:correff}, it can be applied in an \ai context with inclusion of the relevant many-body effects. After detailing how nonlinear optical susceptibility is extracted from the dynamical polarization (Sec.~\ref{sc:compdet}), we show results for the second- and third-harmonic generation (THG) in semiconductors (Sec.~\ref{sc:results}) and successfully validate them against existing results from the literature obtained by response theory in frequency domain.   
   
%The Berry phase was introduced in the context of the , a theory that deals with the definition of static bulk polarization for many-electron correlated systems and noncrystalline systems, by showing that the correct coupling of the electronic wavefunction with the electric field is through a geometric many-body phase operator. 

%% %\textcolor{red}{In the introduction dovremmo discutere anche un po' di non linear optics, citare Sipe, Luppi ed anche Berkaine\cite{PhysRevB.83.245205}, quest'ultimo ha usato la fase di Berry per calcolare la X2 e X3 statica nel TeO2}

\section{Theoretical background}\label{sc:theory}
We consider a system of $N$ electrons in a crystalline solid of volume $V=Mv$ (where $M$ is the number of the equivalent cells and $v$ the cell volume) coupled with a time-dependent electric field $\efield$
\be
H(t)=H^0 + H^{\efield}(t), \label{eq:startH}
\ee
where $H^0$ is the zero-field Hamiltonian, and $H^{\efield}(t)$ describes the coupling with the electric field. % is treated within the dipole approximation ($-e$ is the electronic charge).
%In this section we do not specify the $H^0$ Hamiltonian but consider a generic one particle Hamiltonian that respects Born-von K\'arm\'an periodic boundary condition.\\
%To have a computational treatable problem, $H^0$ in Eq.~\eqref{eq:startH} is usually approximated treating the electron-electron interaction through an effective one-particle operator. 
Here, we consider a generic single-particle Hamiltonian $H^0$. In Sec.~\ref{ss:correff} we specify the form of $H^0$ and show how many-body effects are included by means of effective single-particle operators. Of course, the choice of a single-particle Hamiltonian prevents applications to systems with strong static correlation such as Mott insulators or frustrated magnetic materials.
%that respects Born-von K\'arm\'an periodic boundary condition.\\
We assume the ground state of $H^0$ to be non-degenerate and a spin-singlet so that the ground-state wavefunction can be expressed as a single Slater determinant. 
%We assume that $H^0$ is such that the many-body ground-state wavefunction can be expressed as a single Slater determinant (see Sec.~\ref{ss:correff}). Of course, this choice prevents applications to systems with strong static correlation.  
We also assume, as usual in treating cell-periodic systems, Born-von K\'arm\'an PBC and define a regular grid of  $N_\kk=M$  $\kk$-points in the Brillouin zone. With such assumptions, the single-particle solutions of $H^0$ are Bloch-functions.

Regarding the electron-field coupling we assume classic fields and use the dipole approximation, $H^{\efield}(t)=e\efield(t)\hat r$ ($-e$ is the electronic charge).  However, because of the PBC the position operator is ill-defined. In order to obtain a form for the field coupling operator compatible with  Born-von K\'arm\'an PBC, in this paper we use the Berry phase formulation of the position operator and  consequently the polarization. As proved in Ref.~\onlinecite{souza_prb}, in this formulation the solutions of  $H(t)$ are also in a Bloch function form: $\phi_{\kk,n}(\rr,t) = \mathrm{exp}(i\kk\cdot\rr) v_{\kk,n} (\rr,t)$, with  $v_{\kk,n}$ being the periodic part and $n$ being the band index. Notice that, even in the Berry phase formulation, for very strong fields and with the number of $\kk$-points that goes to infinity the Hamiltonian Eq.~\ref{eq:startH} is unbounded from below due to the Zener tunnelling.\cite{springborg} Nevertheless the strength of the fields used in non-linear optics is well below this limit.\cite{springborg,souza_prb}\\
In Sec.~\ref{ss:fldcpl} we detail how, by starting from the Berry phase formulation of polarization, we  obtain the EOMs in presence of an external electric field within PBC.                 
\subsection{Treatment of the field coupling term}\label{ss:fldcpl}
\subsubsection{Berry phase polarization}
Developed in the mid-90s the Modern Theory of Polarization~\cite{RevModPhys.66.899} provides a correct definition for the macroscopic bulk polarization, not limited to the perturbative regime, in terms of the many-body geometric phase 
\be 
%\langle X \rangle &=& \frac{L}{2\pi}
%\mbox{Im ln}  \langle \Psi_0 | {\rm e}^{i\frac{2\pi}{L} \hat{X}} | \Psi_0
%\rangle \label{main} \\
%\PP_\alpha =  \lim_{V \rightarrow \infty} \frac{e}{2\pi} \mbox{Im ln }  \langle \Psi_0 | {\rm e}^{i \bb_\alpha \cdot \hat{\mathbf X}} | \Psi_0 \rangle . \label{limit} 
\PP_\alpha =  \frac{e N_{\kk_\alpha} \lv_\alpha}{2\pi V} \mbox{Im ln }  \langle \Psi_0 | {\rm e}^{i \qq_\alpha \cdot \hat{\mathbf X}} | \Psi_0 \rangle . \label{limit} 
\ee
In Eq.~\eqref{limit} $\PP_\alpha$ is the macroscopic polarization along the primitive lattice vector $\lv_\alpha$, $\hat{\mathbf X} = \sum_{i=1}^{N} \hat{\mathbf x}_i$, $\qq_\alpha = \frac{\bb_\alpha}{N_{\kk_\alpha}}$ with $\bb_\alpha$ the primitive reciprocal lattice vector such that $\bb_\alpha\cdot\lv_\alpha=2\pi$, and $N_{\kk_\alpha}$ the number of $\kk$-points along $\alpha$, corresponding to the number of equivalent cells in that direction.
Note that in this formulation the polarization operator is a genuine many-body operator that cannot be split as a sum of single-particle operators. 

%% By using the assumption that the full many-body wave-function can be written as a single Slater determinant to simplify Eq.~\eqref{limit} we assume . 

%% A further simplication comes by specializing Eq.~\ref{limit} for crystalline systems, that is by assuming that our system is composed of $n_{\text{cell}}$ equivalent cells, 
%, where $L=N_{\kk} a$ and $a$ is the lattice vector. 

By using the assumption that the wave-function can be written as a single Slater determinant,
the expectation value of the many-body geometric phase in Eq.~\eqref{limit} can be seen as the overlap between two single Slater determinants. % As such it is equal to the determinant of the overlap $\cal S$ matrix built out of  $\phi_{\kk_j,m}$, the occupied Bloch functions
The latter is equal to the determinant of the overlap $\cal S$ matrix built out of $\phi_{\kk_j,m}$, the occupied Bloch functions
\be
{\cal S}_{\kk m, \mathbf{k'} m'} = \langle \phi_{\kk,m} | e^{-i \mathbf q_\alpha \hat x}|\phi_{\mathbf {k'},m'} \rangle \label{sint}.
\ee 
%with $\qq_\alpha = \frac{\bb_\alpha}{N_{\kk_\alpha}}$, where $N_{\kk_\alpha}$ is the number of $\kk$-points along $\bb_\alpha$.
%$ is the smallest vector that connects two $\kk$-points along the $\bb_\alpha$ direction.

Then we can rewrite Eq.~\eqref{limit} as
\be
\mathbf P_\alpha = -\frac{ef \lv_\alpha}{2 \pi N_{\kk_\alpha^\perp} v} \mbox{Im ln det } {\cal S}, \label{eq:Pipa}
\ee
where $f$ is the spin degeneracy, equal to $2$ since we consider here only spin-unpolarized systems, and $N_{\kk_\alpha^\perp}$ is the number of $\kk$-points in the plane perpendicular to reciprocal lattice vector $\bb_\alpha$, with $N_{\kk} = N_{\kk_\alpha^\perp}\times N_{\kk_\alpha}$.

%% Following these assumptions the full many-body wavefunction can be thus written as:
%% \begin{equation}
%%     | \, \Psi_0 \rangle = {\sf A}  \prod_{j=0}^{N_\kk} \frac{1}{v^{n_b}} \frac{1}{\sqrt{(2n_b)!}}|\phi_{\kk_j,1}\bar \phi_{\kk_j,1} ... \phi_{\kk_j,n_b}\bar \phi_{\kk_j,n_b}|   , \label{det}
%% \end{equation} 
%% where $ {\sf A}$ is the antisymmetrizer operator, $v$ the volume of the cell, $n_b$ is the number of doubly occupied bands, and  $\phi_{\kk_j,m}$ are Bloch functions. By substituting Eq.~\eqref{det} in Eq.~\eqref{limit} (see also Ref.~\onlinecite{PhysRevLett.80.1800,KSV1}) we obtain:
%% %and consider the polarization along a direction parallel to one of the primitive reciprocal lattice vectors $\bb_\alpha$: (already in Eq 2)
%% \be
%% \mathbf P_\alpha = -\frac{ef}{2 \pi}  \mbox{Im ln det } {\cal S}, \label{eq:Pipa}
%% \ee
%% with
%% \be
%% {\cal S}_{\kk m, \mathbf{k'} m'} = \langle \phi_{\kk,m} | e^{-i \mathbf q_\alpha \hat x}|\phi_{\mathbf {k'},m'} \rangle \label{sint},
%% \ee 
%%  In Eq.~\eqref{eq:Pipa} 

The overlap $\cal S$ has dimensions $n_b N_\kk\times n_b N_\kk$, where $n_b$ is the number of doubly occupied bands. However, from the properties of the Bloch functions and by imposing they satisfy the so-called ``periodic gauge'' $\phi_{\kk+ \mathbf G} = \phi_{\kk}$, it follows that the integrals in Eq.~\eqref{sint} are different from zero only if $\kk' -\kk = \qq_\alpha$.  Therefore the determinant of $\cal S$ reduces to the product of $N_{\kk_\alpha}$ determinants of overlaps $S$ built out of $v_{\kk,m}$, the periodic part of the occupied Bloch functions:
\be
S_{mn}(\kk , \kk + \qq_\alpha) = \langle v_{\kk,m} | v_{\kk + \qq_\alpha,n} \rangle. 
\label{eq:sovlps}
\ee
This leads to the formula by which we compute the polarization of the system 
 \begin{equation}
     \mathbf P_\alpha = -\frac{ef}{2 \pi v} \frac{\mathbf a_\alpha}{N_{\kk_\alpha^\perp}} \sum_{\kk_\alpha^\perp} \mbox{Im ln} \prod_{i=1}^{N_{\kk_\alpha}-1}\ \mbox{det } S(\kk_i , \kk_i + \mathbf q_\alpha). \label{berryP} 
 \end{equation}
Using matrix properties,~\cite{mcookbook} the logarithm of the matrix determinant can be rewritten as the trace of matrix logarithm, and so Eq.~\eqref{berryP} can be transformed as 
\be 
\mathbf P_\alpha = -\frac{ef}{2 \pi v} \frac{\mathbf a_\alpha}{N_{\kk_\alpha^\perp}} \sum_{\kk_\alpha^\perp} \mbox{Im} \sum_{i=1}^{N_{\kk_\alpha}-1}\ \mbox{tr ln } S(\kk_i , \kk_i + \qq_\alpha) \label{xtrace}, 
\ee
more suitable to derive the EOMs. By taking the thermodynamic limit ($N_\kk \rightarrow \infty$ and $\qq_\alpha \rightarrow 0$ ) of the latter expression one arrives at the King-Smith and Vanderbilt formula for polarization.~\cite{KSV1} 
% \be
%\mathbf P_\alpha =  i\frac{ef}{2\pi} \frac{1}{N_{\kk_\perp}} \sum_{\kk_\perp} \sum_{\kk_\alpha}^{N_{\kk_\alpha}-1} \sum_{m=1}^{M} \langle v_{\kk,m} | \partial_k v_{\kk,m} \rangle + O(\mathbf q^2). \label{mypolarization} 
%\ee
Since in a numerical implementation we deal with a finite number of $N_\kk$ and finite $\qq_\alpha$, we stick here to Eq.~\eqref{xtrace} with $\qq_\alpha = \Delta \kk_\alpha$ to derive the EOMs.  

% \mbox{det} \; S = \prod_{j=0}^{N_{\kk}-1} \mbox{det}\; S(\kk_j,\kk_{j+1}) , 
\subsubsection{Equations-of-motion}
Following Ref.~\onlinecite{souza_prb} we start from the Lagrangian of the system in presence of an external electric field $\efield$:
\be
{\cal L}=\frac{i\hbar}{N_\kk}\sum_{n=1}^M \sum_{\kk}\,
\langle v_{\kk n}|\dot{v}_{\kk n} \rangle-E^0 - v \efield\cdot\PP,
	\label{eq:lagrangian_discrete} 
\ee
where $E^0$ is the energy functional corresponding to the zero-field Hamiltonian:
\be
 E^0= \frac{1}{N_\kk} \sum_{n=1}^M \sum_{\kk}\, \langle v_{\kk n}|\hat H_\kk^0 | v_{\kk n} \rangle, 
\ee
with $\hat H_\kk^0 = e^{-i\kk\rr'}H^0 e^{i\kk\rr} $. Notice that $H^0$ does not connect wave-functions with different $\kk$ vectors.
To simplify the notation we do not explicit the time dependence of the $|v_{\kk n}\rangle$, but they should be considered time-dependent in the rest of the paper.  

We derive the dynamical equations from the Euler-Lagrange equations
\bea
\frac{d}{dt}
\frac{\delta \cal L}{ \langle \delta \dot{v}_{\kk,n} |}-\frac{\delta \cal L}{\langle \delta v_{\kk,n} |} &=&0,\\
i\hbar \frac{d}{dt} |v_{\kk n}\rangle -
\hat H_\kk^0 |v_{\kk n}\rangle -N_\kk v \efield\cdot\frac{\delta \PP}{\langle \delta v_{\kk,n} |} &=&0.
\label{eq:euler-lagrange_b}
\eea
To obtain the functional derivative of the polarization expression in Eq.~\eqref{xtrace} we use that~\cite{souza_prb,gonze} 
\be
\delta\mbox{tr ln} S = \text{tr}\left[S^{-1}\delta S\right] + {\cal O}(\delta S^2), 
\ee
%(see for example Appendix C.2 of Ref.~\onlinecite{souza_prb} or Appendix B of Ref.~\onlinecite{gonze})
and that exchanging arguments ($\kk\leftrightarrow \kk'$) in $S$ [Eq.~\eqref{eq:sovlps}] brings a minus sign in Eq.~\eqref{xtrace}.
This leads to (see Ref.~\onlinecite{souza_prb} for details):
\bea
\frac{\delta \PP_\alpha}{\langle \delta v_{\kk,n}|} &=& -\frac{ief}{2 \pi} \frac{\lv_\alpha}{2 N_{\kk_\alpha^\perp} v} \left( | \tilde v_{\kk^{+}_\alpha,n} \rangle - | \tilde v_{ \kk^{-}_\alpha,n} \rangle\right) \label{eq:fdpol}\\
| \tilde v_{ \kk^{\pm}_\alpha,n} \rangle &=& \sum_{m} \left(S(\kk,\kk^{\pm}_\alpha\right)^{-1})_{mn} |v_{\kk^{\pm}_\alpha,m}\rangle, \label{eq:vtilde} 
\eea
where $\kk^{\pm}_\alpha = \kk \pm \Delta \kk_\alpha$,
%%  \bea
%% \efield |\tilde \partial v_{\kk,n} \rangle &\simeq& \frac{1}{4 \pi} \sum_{i=1}^{3} \frac{ |\bb_i|}{\Delta \kk^\alpha_i} (\efield \cdot \mathbf a_i) \left \{ | \tilde v_{ \kk^+_i,n} \rangle - | \tilde v_{ \Delta \kk^-_i,n} \rangle  \right \} \nonumber \\
%%   | \tilde v_{ \kk^+_i,n} \rangle &=& \sum_{m=1}^M S^{-1}(\kk,\kk + \Delta \kk^\alpha_i)_{m,n}  | v_{ \kk+\Delta \kk^\alpha_i,m} \rangle     \label{covderiv}
%% \eea                                                                                                            
and from which we can define the field coupling operator
\begin{align}
\hat w_\kk(\efield) =\frac{ief}{4 \pi}\sum_m \sum_{\alpha=1}^3 \left(\lv_\alpha\cdot\efield\right)N_{\kk_\alpha} \sum_{\sigma=\pm} \sigma |\tilde v_{\kk^\sigma_\alpha,m} \rangle\langle v_{\kk,m}|. 
\label{eq:fldcpl}
\end{align}
Notice that the field coupling operator in Eq.~\eqref{eq:fldcpl} is nonhermitian. In order to have well defined Hermitian operators in the EOMs we replace $\hat w_\kk(\efield)$ with $\hat w_\kk(\efield) + \hat w^\dagger_\kk(\efield)$. This is possible because at any time $\hat{\rm w}_{\kk}^{\dagger}|v_{\kk n}\rangle=0$.\cite{souza_prb}  
Finally, by using Eqs.~\eqref{eq:fdpol}-\eqref{eq:fldcpl} in Eq.~\eqref{eq:euler-lagrange_b} and the Hermitian field coupling operator we obtain the EOMs:
\be
i\hbar  \frac{d}{dt}| v_{\kk,m} \rangle = \left(\hat H_\kk^0 + \hat w_\kk(\efield) +\hat w^\dagger_\kk(\efield) \right)| v_{\kk,m} \rangle. \label{eom}
\ee

Note that Eq.~\eqref{eq:fldcpl} contains 
\be
\frac{1}{2\Delta\kk_\alpha}\left( | \tilde v_{\kk^{+}_\alpha,n} \rangle - | \tilde v_{ \kk^{-}_\alpha,n} \rangle\right)
\label{eq:ncvdrv}
\ee
that has the form of the two-points central finite difference approximation of $\partial_{\kk_\alpha}|v_{\kk_\alpha}\rangle$, but for the fact that $|\tilde v_{\kk^{\pm}}\rangle$ are used instead of $|v_{\kk^{\pm}}\rangle$. As explained in Ref.~\onlinecite{souza_prb}, the $|\tilde v_{\kk^{\pm}}\rangle$ are built from the $|v_{\kk^{\pm}}\rangle$ [Eq.~\eqref{eq:vtilde}] in such a way that they transform as $|v_\kk\rangle$ under a unitary transformation $U_{\kk,nn'}$. 

In fact, there is a gauge freedom in the definition of $|v_\kk\rangle$,  that is $|v_\kk\rangle \rightarrow  U_{\kk} |v_\kk\rangle$, and since the Hamiltonian is diagonalized independently at each $\kk$, the gauge is fixed independently and randomly at each $\kk$. Then, standard (numerical) differentiation will be affected by the different gauge choices at two neighbouring $\kk$-points. Instead the (numerical) derivative in Eq.~\eqref{eq:ncvdrv} is gauge-invariant, or more specifically is performed in a locally flat coordinate system with respect to $U_{\kk,nn'}$. In fact, in the thermodynamical/continuum limit, Eq.~\eqref{eq:ncvdrv} corresponds to the covariant derivative. The problem of differentiating $|v_\kk\rangle$  with respect to $\kk$ has been addressed also in Refs.~\onlinecite{PhysRevB.76.035213,springborg,andrew1998computation} that use alternative approaches to ensure the gauge-invariance.
In the here-discussed approach the definition of a numerical covariant differentiation originates directly from the definition of the polarization as a Berry phase.

\subsection{Treatment of electron correlation}\label{ss:correff}
%In the previous sections we described a general approach to study real-time response in solids within the independent particle approximation. However is well known, that  
Correlation effects play a crucial role in both linear\cite{RevModPhys.74.601} and non linear\cite{PhysRevB.82.235201,PhysRevB.80.155205} response of solids. %% To simplify Eq.~\eqref{limit} we restricted the choice of $\HH^0_\kk$, thus of possible correlation effects, so that the system wavefunction $|\Psi_0\rangle$ can be written as a single Slater determinant. 
Since we assumed that $|\Psi_0\rangle$ in Eq.~\eqref{limit} can be written as a single Slater determinant, effects beyond the IPA can be introduced in $\hat H^0$ through an effective time-independent one-particle operator that can be either spatially local as in time-dependent density functional theory, or spatially nonlocal as in time-dependent Hartree-Fock. 

However, both time-dependent density functional theory and time-dependent Hartree-Fock are not suitable approaches to optical properties of semiconductors: the former, within standard approximations for the exchange-correlation approximations, underestimates the optical gap and misses the excitonic resonances; the latter largely overestimates the band-gap and excitonic effects.   

In the framework of Green's function theory a very successful way to deal with electron-electron interaction in semiconductors is the combination of the $G_0W_0$ approximation for the quasi-particle band structure~\cite{PhysRevB.25.2867} with the Bethe-Salpeter equation in static ladder approximation for the response function.~\cite{strinati}  

We recently extended this approach to the real-time domain~\cite{attaccalite} by mean of non-equilibrium Green's function theory. In practice, the latter approach corresponds to a time-dependent static screened Hartree-Fock operator that satisfies the above-mentioned restrictions on the choice of $\hat H^0$ and thus can be used within the here proposed framework. In what follows, we reformulate the approach in Ref.~\onlinecite{attaccalite} as time-dependent Schr\"odinger like equations [Eq.~\eqref{eom}]. 

As starting point we choose the Kohn-Sham  Harmiltonian at fixed density as a system of indipendent particles,~\cite{PhysRev.140.A1133} 
\be
\hat H^{0,\text{IPA}} \equiv \hat h^{\text{KS}} = -\frac{\hbar^2}{2m}\sum_{i} \nabla_i^2 + \hat V_{eI} + \hat V_{H}[\rho^0]+ \hat V_{\text{xc}}[\rho^0],      
\label{eq:HIPA}
\ee
where $V_{eI}$ is the electron-ion interaction, $V_{H}$ the Hartree potential and $V_{\text{xc}}$ the exchange-correlation potential.
The advantage of such a choice is that the Kohn-Sham system is the independent-particle system that reproduces the electronic density of the unperturbed many-body interacting system $\rho^0$, thus by virtue of the Hohenberg-Kohn theorem~\cite{PhysRev.136.B864} the ground-state properties of the system. Furthermore, no material dependent parameters need to be input, but for the atomic structure and composition. 

As first step beyond the IPA, we introduce the corrections to the independent-particle energy levels by the electron-electron interaction through a (state-dependent) scissor operator 
\be
\Delta \hat H = \sum_{n,\kk} \Delta_{n,\kk} |v^0_{n,\kk}\rangle\langle v^0_{n,\kk}|.
\ee
 The latter can be calculated \ai e.g., via the $G_0W_0$ approach $\Delta_{n,\kk} = (E^{G_0W_0}_{n,\kk} - \varepsilon^{\text{KS}}_{n,\kk} ) $, or can be determined empirically from the experimental band gap  $\Delta_{n,\kk} = \Delta = E^{\text{exp}}_{\text{GAP}} - \Delta\varepsilon^{\text{KS}}_{\text{GAP}}$. We refer to this approximation as the independent quasiparticle approximation (QPA): 
\be
\hat H^{0,\text{QPA}} \equiv \hat h^{\text{KS}} + \Delta \hat H. 
\label{eq-tdqpa}
\ee
Notice that in our approach the inclusion of a non-local operator in the Hamiltonian does not present more difficulties than a local one, while in the response theory in frequency domain this is not a trivial task.\cite{PhysRevB.82.235201} 
As a second step we consider the effects originating from the response of the effective potential to density fluctuations. By considering the change of the Hartree plus the exchange-correlation potential in Eq.~\ref{eq:HIPA} we will obtain the TD-DFT response. Here we include just ``classic electrostatic'' effects via the Hartree part. We refer to this level of approximation as the time-dependent Hartree (TDH)
\be
\hat H^{0,\text{TDH}} \equiv \hat H^{0,\text{QPA}} + \hat V_{H} [\rho-\rho^0]. 
\label{eq-tdh}
\ee
In the linear response limit the TDH is usually referred as Random-Phase approximation and is responsible of the so-called crystal local field effects.\cite{PhysRev.126.413} 

Beyond the TDH approximation one has the TD-Hartree-Fock that includes the response of the exchange term to fluctuations of the density matrix $\gamma$. As discussed above this level of approximation is insufficient for optical properties of semiconductors, normally worsening over TDH results. The next step is thus to consider a screened exchange term in which the relevant electron correlation is introduced as a static screening term.~\cite{strinati} The latter is calculated for the unperturbed KS system and is fixed to its initial value.
We refer to this level of approximation as TD screened exchange or TD screened Hartree-Fock (TD-SHF),
\be
\hat H^{0,\text{TDSHF}} \equiv \hat H^{0,\text{TDH}} + \hat \Sigma^{0,\text{SHF}}[\gamma - \gamma^0].
\ee
We want to emphasize again that within this approach many-body effects are easily implemented by adding terms to the unperturbed independent-particle Hamiltonian $\hat H^{0,\text{IPA}}$ in the EOMs [Eq.~\eqref{eom}]. 
Limitations may arise because of the computational cost of calculating those addition terms. In the specific the large number of $\kk$-points needed to converge the SHG and THG spectra makes TD-SHF calculations impracticable. However, much less $\kk$-points are needed for converging the screened-exchange self-energy itself and currently we are investigating how to exploit this property and devise ``double grid'' strategy similar to the one proposed in Ref.~\onlinecite{kammerlander}. In this work effects beyond IPA are limited to the QPA and TDH.

Finally, when the wave-function cannot be approximated anymore with a single Slater determinant (as in strong-correlated systems) the evaluation of the polarization operator [Eq.~\ref{limit}] becomes quite cumbersome.\cite{stella} Also we are not aware of any successful attempt to combine Berry's phase polarization with Green's function theory or density matrix kinetic equations (including for example scattering terms), even if some appealing approaches have been proposed.\cite{restagw,PhysRevB.84.205137}

\section{Computational scheme and numerical parameters}\label{sc:compdet}

\begin{figure}[ht]
\centering
\epsfig{figure=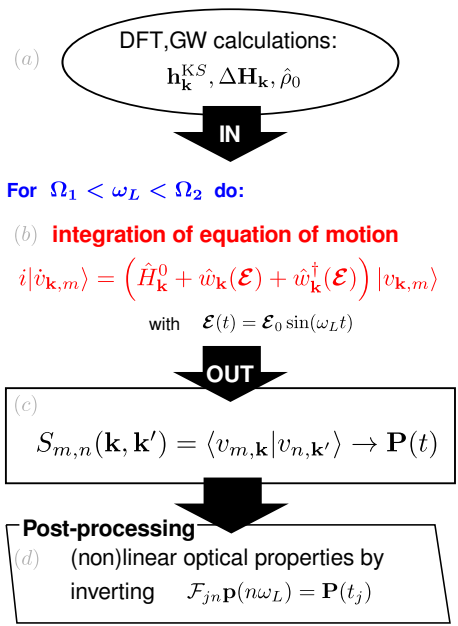,width=6cm,clip}
\caption{\footnotesize{[Color online] Proposed real-time \ai scheme to compute SHG and THG spectra in the $\[\Omega_1,\Omega_2\]$ energy range for extended systems with PBC: $(a)$ Results from KS-DFT and $G_0W_0$ are input to determine the zero-field Hamiltonian. $(b)$ The EOMs [Eq.~\eqref{eom}] are then integrated to obtain $(c)$ the overlaps $S$ from which the polarization is computed as in Eq.~\eqref{berryP}. In the post-processing step $(d)$ the nonlinear susceptibilities are obtained by inversion of the Fourier matrix [Eq.~\eqref{eq:fouinv}], see Sec.~\ref{sc:compdet} for details.}}
\label{fg:cmpscm}
\end{figure}

Figure~\ref{fg:cmpscm} illustrates the computational scheme we use to calculate the SHG and THG spectra. It consists in KS-DFT and $G_0W_0$ calculations to determine the density and the KS eigenvalues, the quasiparticle corrections and eigenfunctions entering the zero-field Hamiltonian; the integration of the equation of motions [Eq.~\eqref{eom}] with a monochromatic electric field $\efield(t) = \efield_0 \sin(\omega_L t)$ to obtain the $\PP(t)$ from Eq.~\eqref{berryP} and the post-processing of $\PP(t)$ to extract the nonlinear susceptibilities. The latter two steps are repeated varying the laser frequency $\omega_L$ within the energy range for which we calculate the spectra. 

The scheme in Fig.~\ref{fg:cmpscm} has been implemented in the development version of the {\sc Yambo} code.~\cite{yambo} 
Kohn-Sham calculations have been performed using the {\sc Abinit} code,~\cite{abinit} and the relevant numerical parameters are summarized in Table~\ref{tb:cmppar}. All the operators  appearing in the EOMs[Eqs.~\eqref{eom},\eqref{eq-tdh},\eqref{eq-tdqpa}] have been expanded in the Kohn-Sham basis set and the number of bands employed in the expansion is reported in Table~\ref{tb:cmppar}. 

Rigorously to have a fully \ai scheme, the scissor operator has to be calculated using e.g., $G^0W^0$. Here we use an empirical values for the scissor operator (reported in Table.~\ref{tb:cmppar}) since the scope is to validate the computational scheme, and to facilitate the comparison with other works in the literature.

\begin{table}[hb]
  \begin{tabular}{l|c|c|c|c|c|c}
    System & PP & $E_{\text{c}}$(Ha) & $\kk$-grid & $a$ (\AA)& Bands &$\Delta$(eV)\\ 
      \hline
      \hline
    SiC &  Si:$(3s)^2(3p)^2$    & 30 & 8/16 &4.36 &1-8 & 0.8\\
        & C:$(2s)^2(2p)^2$     &&&&\\
      \hline
    AlAs& Al:$(3s)^2(3p)^1$    & 20 & 8/18 &5.66 & 2-10 & 0.9 \\
        & As:$(4s)^2(4p)^3$     &&&&\\
      \hline
    CdTe& Cd:$(4d)^{10}(5s)^2$    & 40 & 8/14&6.48&7-13&1.0 \\
        & Te:$(5s)^2(5p)^4$    &  & &  &  \\
      \hline
    Si  & Si:$(3s)^2(3p)^2$ & 14 & 8/24&5.39&1-7&0.6 \\\hline
      \hline
  \end{tabular}
  \caption{\footnotesize{Parameters used in the Kohn-Sham density-functional theory and in the real-time simulations: the pseudopotential components for each atom (PP); energy cutoff for the planewaves ($E_{\text{c}}$); number of $\kk$ points of the Monkhorst-Pack grid in each of the three dimensions for calculating the density and in the RT-simulations ($\kk$-grid); the lattice parameter ($a$); the range of bands for which the single-particle wave-function is evolved during the RT-simulation; the QP scissor ($\Delta$) used within the QPA.}}
  \label{tb:cmppar}
\end{table}

The EOMs [Eq.~\eqref{eom}] have been integrated using the following algorithm~\cite{koonin90}
\be
\label{eq:time_evolution}
\ket{v_{\kk n}(t+\Delta t)}= \frac{I-i(\Delta t/2)\hat H^0_{\kk}(t)}{I+i(\Delta t/2) \hat H^0_{\kk}(t)} \ket{v_{\kk n}(t)},
\ee
valid for both Hermitian and non-Hermitian Hamiltonians, and strictly unitary for any value of the time-step $\Delta t$ in the Hermitian case. In all real-time simulations we used a time-step of 0.01 fs. The number of states (number of bands $\times$ number of $\kk$ points) evolved during the simulations is reported in Table~\ref{tb:cmppar}. \\

\begin{figure}[hb]
\centering
\epsfig{figure=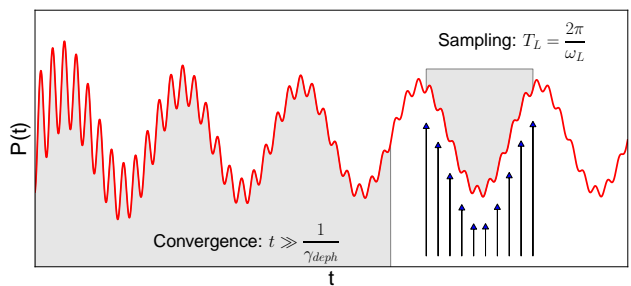,width=8.5cm,clip}
\caption{\footnotesize{[Color online] Pictorial representation of the signal analysis in the post-processing step. The signal  $P(t)$ (red line) can be divided into two regions: an initial convergence region (up to $t\gg 1/\gamma_{deph}$) in which the eigenfrequencies of the systems are ``filtered out'' by dephasing and a second region where Eq.~\eqref{eq:frrexp} holds. In this second region the signal $P(t)$ is sampled within a period $T_L=2\pi/\omega_L$ to extract the $P^\alpha_i$ coefficients of Eq.~\ref{eq:fouinv}. Note that $P(t)$ is not a realistic one: for illustration purposes we enhanced the second-harmonic signal that otherwise would not be visible on this scale.}} 
\label{fg:ptanalysis}
\end{figure} 

In our simulations we switch on the monochromatic field at $t=t_0$. This sudden switch excites the eigenfrequencies of the system $\omega^{0}_l$ introducing spurious contributions to the non-linear response.
We thus add an imaginary term into the Hamiltonian $H^{0}_\kk$ to simulate a finite dephasing:   
\be
\Gamma= -\frac{i}{\gamma_{\text{deph}}} \sum_l  \{  | v_{\kk,l}\rangle \langle  v_{\kk,l} | - | v^0_{\kk,l}\rangle \langle  v^0_{\kk,l} | \}
\ee
where $|v^0_{\kk,l}\rangle$ are the valence bands of the unperturbed system and $\gamma_{\text{deph}}$ is the dephasing rate. Then we run the simulations for a time much larger than $1/\gamma_{\text{deph}}$ and sample $\PP(t)$ close to the end of the simulation, see Figure~\ref{fg:ptanalysis}.
Since $\gamma_{\text{deph}}$ determines also the spectral broadening, we cannot choose it arbitrary small. For example in the present calculations we have chosen $1/\gamma_{\text{deph}}$ of 6 fs that corresponds to a broadening of approximately 0.2 eV (comparable with the experimental one) and thus we run the simulations for 50-55 fs.\\
Once all the eigenfrequencies of the system are filtered out, the remaining polarization $\PP(t)$ is a periodic function of period $T_L =\frac{2\pi}{\omega_L}$, where $\omega_L$ is the frequency of the external perturbation and can be expanded in a Fourier series
\be\label{eq:frrexp}
\PP(t) = \sum_{n=-\infty}^{+\infty} \pp_n e^{-i\omega_n t},
\ee  
with $\omega_n = n \omega_L$, and complex coefficients:
\begin{equation}\label{eq:frrcff}
 \pp_n = \mathscr{F}\{\PP(\omega_n)\} =\int_{0}^{T_L} dt \PP(t) e^{i\omega_n t}.
\end{equation}
To obtain the optical susceptibilities of order $n$ at frequency $\omega_L$ one needs to calculate the $\pp_n$ of Eq.~\eqref{eq:frrexp}, proportional to $\susc n$ by the $n$-th power of the $\efield_0$. 
However, the expression in Eq.~\eqref{eq:frrcff} is not the most computationally convenient since one needs a very short time step---significantly shorter than the one needed to integrate the EOMs---to perform the integration with sufficient accuracy. As an alternative we use directly Eq.~\eqref{eq:frrexp}: we truncate the Fourier series to an order $S$ larger than the one of the response function we are interested in. We sample $2S+1$ values $\PP_i\equiv\PP(t_i)$ within a period $T_L$, as illustrated in Figure~\ref{fg:ptanalysis}. Then Eq.~\eqref{eq:frrexp} reads as a system of linear equations 
\be
{\cal F}_{in} p^\alpha_n = P^\alpha_i,
\label{eq:fouinv}
\ee 
from which the component $p^\alpha_n$ of $\pp_n$ in the $\alpha$ direction is found by inversion of the $(2S+1)\times(2S+1)$ Fourier matrix ${\cal F}_{in} \equiv \exp(-i\omega_n t_i)$. We found that the second harmonic generation converges with S equal to 4 while the third harmonic requires S equal to 6. Finally we noticed that averaging averaging the results on more periods can slightly reduce the numerical error in the signal analysis. \\
Alternatively one can opt for a slow switch on of the electric field as in Takimoto et al.,\cite{takimoto:154114} so that no eigenfrequencies of the system are excited, and avoid to introduce imaginary terms in the Hamiltonian. We found, however, that the latter approach also requires long simulations, and on the other hand, it is less straightforward to extract the $\susc n$.

\section{Results}\label{sc:results}
The main objective of this section is to validate the computational approach described in Secs.~\ref{sc:theory} and \ref{sc:compdet} against results in the literature for SHG obtained by the response theory in frequency domain. In particular we chose to validate against results from Refs.~\onlinecite{PhysRevB.82.235201,PSSB.427.1984} on bulk SiC and AlAs in which the electronic structures is obtained---as in our case---from a pseudopotential plane-wave implementation of Kohn-Sham DFT with the local density approximation, which makes the comparison easier.
In the following we considered the zinc-blende structure of SiC and AlAs for which the $\susc 2$ tensor has only one independent nonzero component, $\susc 2_{xyz}$ (or its equivalent by permutation).

\begin{figure}[ht]
\centering
\epsfig{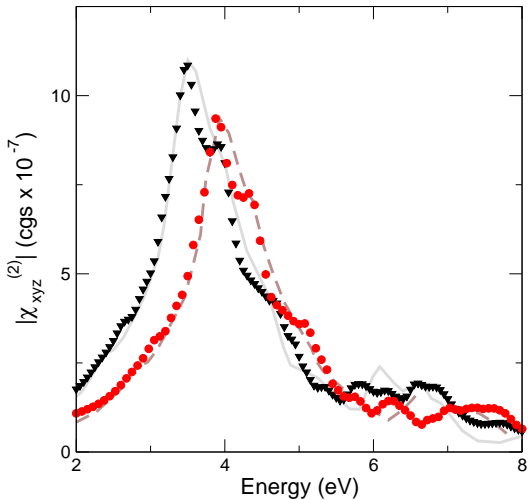}
\caption{\footnotesize{[Color online] Magnitude of $\chi^{(2)}(-2\omega,\omega,\omega)$ for bulk SiC calculated within the IPA (black triangles) and QPA (red circles). Each point corresponds to a real-time simulation at the given laser frequency (see Sec.~\ref{sc:compdet}). Comparison is made with results obtained \ai by direct evaluation of the $\chi^{(2)}$ in Ref.~\onlinecite{PhysRevB.82.235201} in IPA (grey solid line) and QPA (brown dashed line).  \label{fg:SiCQPA} }}
\end{figure}

\begin{figure}[ht]
\centering
\epsfig{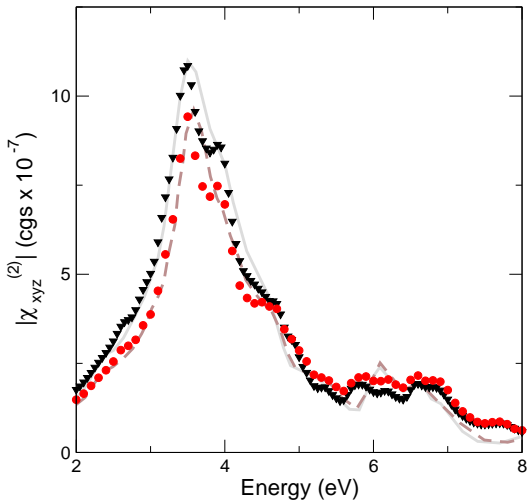}
\caption{\footnotesize{[Color online] Magnitude of $\chi^{(2)}(-2\omega,\omega,\omega)$ for bulk SiC calculated within the IPA (black triangles) and TDH (red circles). Each point corresponds to a real-time simulation at the given laser frequency (see Sec.~\ref{sc:compdet}). Comparison is made with results obtained \ai by direct evaluation of the $\chi^{(2)}$ in Ref.~\onlinecite{PhysRevB.82.235201} in IPA (grey solid line) and TDH (brown dashed line).  \label{fg:SiCRPA} }}
\end{figure}

\begin{figure}[h]
\centering
\epsfig{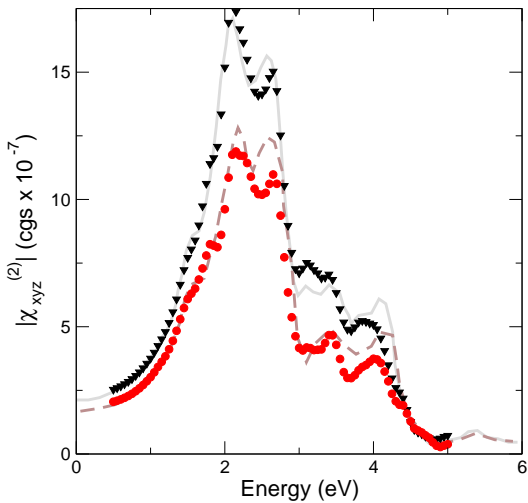}
\caption{\footnotesize{[Color online] Magnitude of $\chi^{(2)}(-2\omega,\omega,\omega)$ for bulk AlAs calculated within the IPA (black triangles) and TDH (red circles). Each point corresponds to a real-time simulation at the given laser frequency (see Sec.~\ref{sc:compdet}). Comparison is made with results obtained \ai by direct evaluation of the $\chi^{(2)}$ in Ref.~\onlinecite{PhysRevB.82.235201} in IPA (grey solid line) and TDH (brown dashed line).  \label{fg:AlAsRPA} }}
\end{figure}

Figures~\ref{fg:SiCQPA} and \ref{fg:SiCRPA} show results for the magnitude of SHG in SiC at the IPA, QPA and TDH level of theory. 
At all levels of approximation we obtained an excellent agreement with the results in Ref.~\onlinecite{PhysRevB.82.235201}. The minor discrepancies between the curves are due to the different choice for the $\kk$-grid used for integration in momentum space: we used a $\Gamma$-centered uniform grid (for which we can implement the numerical derivative) whereas Ref.~\onlinecite{PhysRevB.82.235201} used a shifted grid. Figures~\ref{fg:AlAsQPA} and \ref{fg:AlAsRPA} show results for the magnitude of SHG in AlAs at the IPA, QPA and TDH level of theory. 
Also in this case results obtained from our real-time simulations agree very well with the reference results and again the small differences between the spectra can be ascribed mostly to the different grid for $\kk$-integrations.

As side results we can also observe the effects of different levels of approximation for the Hamiltonian on the SHG spectrum. In order to interpret those spectra note that SHG resonances occur when either $\omega_L$ or $2\omega_L$ equals the difference between two single-particle energies. Then one can distinguish two energy region: below the single-particle minimum direct gap where only resonances at $2\omega_L$ can occur, and above where both $\omega_L$ or $2\omega_L$ resonances  can occur. 
%DESCRIBE IPA for SiC/AlAs?

Regarding the quasiparticle corrections to the IPA energy levels by a scissor operator, below the minimum Kohn-Sham direct band gap the IPA spectrum is shifted by half of the value of the scissor shift (0.4 eV for SiC and 0.45 eV for AlAs) and the spectral intensity reduced by a factor 1.18 (SiC) and  1.25 (AlAs). Above the minimum Kohn-Sham direct band gap instead the QPA spectrum cannot be simply obtained by shifting and renormalizing the IPA one because of the occurrence of resonances at $\omega_{L}$, that are shifted and renormalized differently.  

Regarding the crystal local field, their global effect is to reduce the intensity with respect to the IPA. For SiC, the intensity is reduced by about 15\% below the gap, while above the band gap TDH and IPA have similar intensities. For AlAs we observe a reduction of about 30\% in intensity for the whole range of considered frequencies, but for frequencies larger than 4 eV (that is where the $\omega_L$ resonances with the main optical transition occur) for which again the TDH and IPA have similar intensities.

\begin{figure}[h]
\centering
\epsfig{figure=AlAs_absX2_QP_vs_Luppi.eps,width=7cm,clip}
\caption{\footnotesize{[Color online] Magnitude of $\chi^{(2)}(-2\omega,\omega,\omega)$ for bulk AlAs calculated within the IPA (black triangles) and QPA (red circles). Each point corresponds to a real-time simulation at the given laser frequency (see Sec.~\ref{sc:compdet}). Comparison is made with results obtained \ai by direct evaluation of the $\chi^{(2)}$ in Refs.~\onlinecite{PhysRevB.82.235201,PSSB.427.1984} in IPA (grey solid and dot-dashed line) and QPA (brown dashed and dotted line).}}
\label{fg:AlAsQPA}
\end{figure}

\begin{figure}[h]
\centering
\epsfig{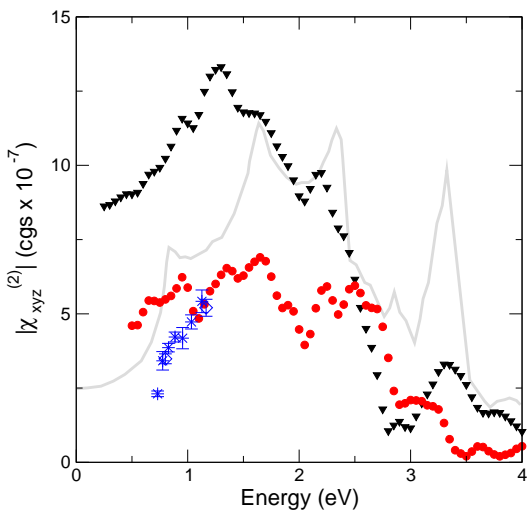}
\caption{\footnotesize{[Color online]  Magnitude of $\chi^{(2)}(-2\omega,\omega,\omega)$ for bulk CdTe calculated within the QPA (black triangles) and TDH (red circles). Each point corresponds to a real-time simulation at the given laser frequency (see Sec.~\ref{sc:compdet}). Comparison is made with results obtained \ai by direct evaluation of the $\susc 2$ in Ref.~\onlinecite{PhysRevB.43.9700} in IPA (grey solid line) and available experimental results in Refs.~\onlinecite{Shoji:97,Jang:13} (blue stars and diamonds).  \label{fg:CdTeSHG} }}
\end{figure}

We also computed the SHG of bulk CdTe (zincblende structure) in Fig.~\ref{fg:CdTeSHG} and we compared with theoretical results~\cite{PhysRevB.43.9700} obtained by a minimal-basis semi-\ai approach (linear combination of Gaussian orbitals in conjunction with an $\alpha$ Slater potential where $\alpha$ is tuned to fit the gap) and with experimental results.~\cite{Shoji:97,Jang:13} At the QPA level (scissor operator of 1.0 eV) the calculated spectrum differs noticeably from the theoretical results in Ref.~\onlinecite{PhysRevB.43.9700}, and largely overestimates the experimental intensities. Interestingly crystal local fields are strong, reducing the intensity of about 50\% with respect to the IPA. The intensity of the TDH spectrum is then consistent with the experimental measurements.

\begin{figure}[ht]
\centering
\epsfig{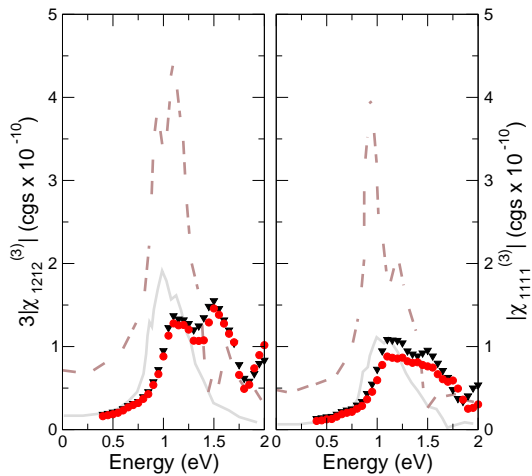}
\caption{\footnotesize{[Color online] Magnitude of the two independent components of the THG in bulk Si: $B=3\chi^{(3)}_{1212}$ (left panel) and $A=\chi^{(3)}_{1111}$ (right panel) calculated within the QPA (black triangles). Each point corresponds to a real-time simulation at the given laser frequency (see Sec.~\ref{sc:compdet}). Comparison is made with results obtained by direct evaluation of the $\chi^{(3)}$ in Ref.~\onlinecite{PhysRevB.41.1542} from empirical (grey solid line) and semi-\ai tight-binding (brown dashed line).  \label{fg:SiX3AB} }}
\end{figure}

\begin{figure}[ht]
\centering
\epsfig{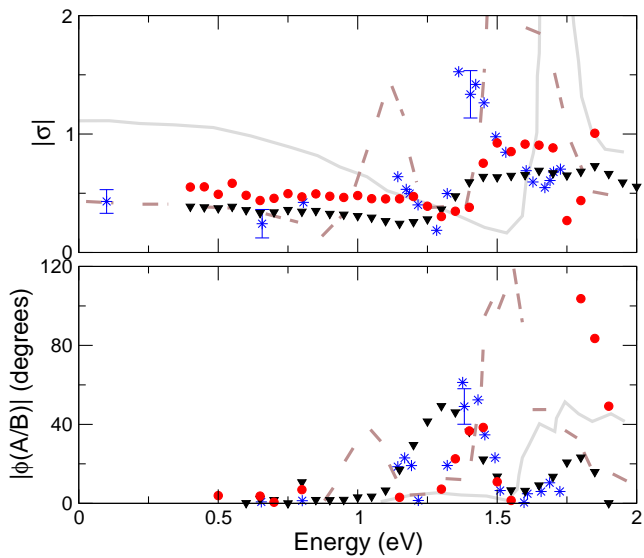}
\caption{\footnotesize{[Color online] Dispersion in the magnitude of the anisotropy $|\sigma| =|(B-A)/A|$ (top panel) and  in the relative phase of $B/A$ (bottom panel) calculated within the QPA (black triangles). Each point corresponds to a real-time simulation at the given laser frequency (see Sec.~\ref{sc:compdet}). Comparison is made with experimental results from Ref.~\onlinecite{moss:89} (blue stars) and results obtained by direct evaluation of the $\chi^{(3)}$ in Ref.~\onlinecite{PhysRevB.41.1542} from empirical (grey solid line) and semi-\ai tight-binding (brown dashed line).  \label{fg:SiX3PA} }}
\end{figure}

Finally, our approach can also compute third order susceptibilities as shown in Figs.~\ref{fg:SiX3AB} and \ref{fg:SiX3PA} for the THG of Si. Within the dipole approximation bulk Si does not have SHG since is centrosymmetric. The first nonlinear effects we can extract from our simulation are at the third order. The THG for Si (diamond structure) has two independent components, $\chi^{(3)}_{1212}  \equiv \chi^{(3)}_{xyxy}$ and $\chi^{(3)}_{1111} \equiv \chi^{(3)}_{xxxx}$.  In the expression for the TH polarization along the direction $i$,
$$P_i(3\omega) = 3\chi^{(3)}_{1212} \efield_i(\omega)|\efield(\omega)|^2+(\chi^{(3)}_{1111} - 3\chi^{(3)}_{1212}) \efield_i^3(\omega),$$
$B=3\chi^{(3)}_{1212}$ is the isotropic contribution, while $A=\chi^{(3)}_{1111}$ the anisotropic contribution. Combination of measurements of THG at different field polarizations provide the anisotropy  $|\sigma| =|(B-A)/A|$ and the phase of $B/A$. Fig.~\ref{fg:SiX3AB} shows our results within the QPA and RPA (scissor operator of 0.6 eV) for $A$ and $B$ compared with theoretical calculations at the tight-binding level with either semi-\ai or empirical parameters. Apparently the  empirical tight-binding results are the closer to our QPA spectra; however, the  semi-\ai tight-binding spectra show the same peak structure and similar ratio between $A$ and $B$ intensities. Both spectra at the RPA level are very similar to the QPA ones: the isotropic contribution is practically identical, while slightly more pronounced differences can be observed for the anisotropic contribution.    
Fig.~\ref{fg:SiX3PA} shows the anisotropy and the phase compared with both other theoretical results and experiments. For energies below 1 eV our QPA spectra is in good agreement with results obtained from  semi-\ai  tight-binding and with the experimental measurement. For higher energies our spectra are less structured with respect both the semi-\ai tight-binding and the experiment. In particular the peak at 1-1.1 eV is missing. On the other hand, the intensities of the spectra  are in better agreement with the experiment than the previous theoretical results. It is interesting to observe how small changes in the spectrum of the anisotropic contribution $A$ due to crystal local field effects induce quite important changes in the phase and anisotropy spectra. In particular, as previously observed they increase the anisotropy. Then, it seems important to include crystal local field effects even if they are weak.

\section{Conclusions}\label{conclusion}                                        
We presented an \ai real-time approach to calculate nonlinear optical properties of extended systems. The key strengths of the proposed approach are first, the correct treatment of the coupling between electrons and external field and second the possibility of easily include effects beyond the IPA.

Regarding the treatment of the electron-field coupling, following the work of Souza et al.\cite{souza_prb}, we started from the Berry-phase formulation for the dynamical polarization---a definition consistent with the PBC---to derive a covariant numerical expression for the dipole operator in the EOMs.

Note that we worked in the length-gauge even if the velocity gauge may appear a more natural choice. In fact, as opposed to the position operator the velocity operator is consistent with the PBC. However, in the velocity gauge even if the position operator disappears from the Hamiltonian, it reappears in the phase factor for the wavefunction,~\cite{PhysRevA.36.2763} so that the problem of re-defining the position operator remains. 
Furthermore, the velocity gauge is plagued by unphysical numerical divergences for the response to low frequencies.~\cite{PhysRevB.52.14636} 
%a fact that has been often forgotten in recent publications which will not be indentified here.
%% Finally we want underline that an alternative approach to study non-linear linear response, the vector potential approach(VPA), has been proposed more than 40 years ago by Genkin and Mednis\cite{Genkin} and recently implemented in modern \ai codes by Kirtman et al.\cite{kirtman:1294,springborg,bishop:7633}. This approach, based on the velocity gauge, keeps correctly into account the wave-function phase factor, and it has been show to be equivalent to the MTP.\cite{springborg} %Notice that in the velocity gauge while the position operator $\mathbf r$ disapears from the Hamiltonian, it reapear as phase factor for the wave-function.\cite{kirtman:1294} This fact has been often forgotten in recent implementation of the velocity gauge\cite{yabana}, however for linear responde this phase factor is irrelevant.\cite{PhysRevA.36.2763}\textcolor{red}{quest'ultima frase possiamo anche toglierla}

Regarding effects beyond the independent-particle approximation, they are included by simply adding the corresponding operator to the single-particle Hamiltonian. This is an easy task when compared with deriving the corresponding expressions for the nonlinear optical susceptibility.~\cite{PhysRevB.80.155205,PhysRevB.80.165318} As an example, in the present work we have included quasiparticle corrections to the band-gap by adding to the Hamiltonian a scissor operator and crystal local-field effects by adding the time-evolution of the Hartree potential. In principle, one can add as well excitonic effects by adding the time-evolution of the screened exchange self-energy as in the scheme proposed in Ref.~\onlinecite{attaccalite}; or perform a real-time TD-DFT calculations by adding the time-evolution of the exchange-correlation potential. Being the focus of this work the validation of the proposed approach for calculating nonlinear properties, we leave the inclusion of these correlation effects for future work.

We have proved the validity of our approach for the SHG in bulk SiC and AlAs by showing an excellent agreement between our results, obtained from real-time simulations, and results in the literature obtained from direct evaluation of the second order susceptibility in frequency-domain.  For CdTe we have computed the SHG and shown that local field effects are important to reproduce experimental measurements. Finally, our approach is not limited to the SHG and we have computed the phase and anisotropy of the THG in bulk Si and obtained results consistent with existing experimental measurements.
  
\section{Acknowledgements} 
The authors acknowledge Angel Rubio, Ivo Souza, Raffaele Resta, and Andrea Marini for useful discussion, and suggestions. M.G. acknowledges the Portuguese Foundation for Science and Technology for funding (PTDC/FIS/103587/2008) and support through the Ci\^encia 2008 program. 
Computing time has been provided by the national GENGI-IDRIS supercomputing centers at Orsay under contract $n^o$ i2012096655 and by the HECToR Supercomputing Facility through EPSRC grant EP/K013459/1, allocated to the UKCP Consortium.

\addcontentsline{toc}{chapter}{Bibliography}
\bibliographystyle{apsrev4-1}
\bibliography{nloptics}
\end{document}